\documentclass[11pt]{article}
\usepackage{amsmath} 
\usepackage{mathtools} 
\usepackage{geometry} 
\begin{document}
\title{Exact density matrix of a discrete quantum system immersed in a thermal reservoir}  

\author{A.J. van Wonderen and L.G. Suttorp\footnote{email: l.g.suttorp@uva.nl} \\ 
Institute of Physics, University of  
Amsterdam, \\ Science Park 904, NL-1098 XH Amsterdam, The Netherlands } 

\maketitle

\begin{abstract} 
Quantum dissipation is studied for a discrete system that linearly interacts with a reservoir of harmonic oscillators at thermal equilibrium. 
Initial correlations between system and reservoir are assumed to be absent. The dissipative dynamics as determined by the unitary evolution 
of system and reservoir is described by a Kraus map consisting of an infinite number of matrices. For all Laplace-transformed Kraus matrices 
exact solutions are constructed in terms of continued fractions that depend on the pair correlation functions of the reservoir. 
\end{abstract}   

{\large

\section{Introduction}  

Recent years have witnessed a lot of research effort on non-Markovian dynamics of open quantum systems. 
Chru\'{s}ci\'{n}ski and Kossakowski \cite{CHR:2013}, Semin and Petruccione \cite{SEM:2016}, as well as 
Ferialdi \cite{FER:2017} surmised that it might be very complicated to derive an exact non-Markovian 
master equation for the evolution of an open quantum system. In this manuscript we obtain an exact 
description of the non-Markovian evolution of an open quantum system. Rather than on a master equation, 
our description is based on a continued-fraction representation of the Kraus map for the density operator 
of the open quantum system.    

In section 2 we derive a finite-temperature description of non-Markovian quantum dynamics on the basis of  
an infinite set of Kraus matrices \cite{KRA:1971}. Our starting-point is the dissipative map for the density 
operator that was obtained in \cite{EPL:2013}. Next, by embedding the set of Kraus matrices in a larger set of 
so-called matrix ratios and performing Laplace transformation, a closed hierarchy of nonlinear equations can 
be constructed, as has been shown in \cite{JPA:2018}. 

In section 3 we iterate the closed hierarchy for matrix ratios an arbitrary but finite number of times. 
Execution of this iteration happens in two steps, the first of which is carried out with relative ease. 
In contrast, the second step is technically most demanding. It provides us with a complicated expression for 
the matrix ratios in terms of matrix continued fractions. 

The identity that is obtained in section 3 contains a matrix ratio of iterative order $N$. In section 4 it is 
assumed that in the limit of $N\rightarrow \infty$ this matrix ratio will be vanishing. By carrying out the 
limit of $N\rightarrow \infty$ we thus obtain a continued-fraction solution for the matrix ratios in closed 
form. Since the Kraus matrices belong to the set of matrix ratios, we obtain the exact solution for the 
density operator of the open quantum system.  

Throughout our paper the following assumptions are made: (i) the Hilbert space of the open quantum system is 
separable, i.e., it is spanned by a countable number of ket vectors; (ii) initial correlations between the 
open quantum system and the surrounding reservoir are absent, so that the initial composite density operator 
for system and reservoir factorizes; (iii) the reservoir consists of a continuum of harmonic oscillators that 
are initially at thermal equilibrium and that linearly interact with the system potentials.  

In addition to the setting (i)-(iii), we furthermore assume that all of the infinite continued fractions 
occurring in this paper are convergent. We recall that the use of continued fractions in quantum optics goes 
back to early work on the Rabi model by Schweber 
\cite{SCH:1967} and Swain \cite{SWA:1973}. 

In summary, our manuscript presents an alternative to master equations, namely an exact representation of the 
Kraus map for an open quantum system in terms of matrix continued fractions.

\section{Kraus hierarchy} 
The evolution in time $t$ of the density matrix $\rho_S (t)$ of an open quantum system $S$
with a discrete energy spectrum is governed by the map \cite{EPL:2013}
\begin{equation} 
\rho_S(t) = {\cal T} \mbox{e}^{L(t)}\rho_S \, , 
\label{1} 
\end{equation}   
where $\rho_S = \rho_S (t=0)$ denotes the initial state of $S$. The superoperator $L$ is
defined as
\begin{eqnarray} 
L(t)\rho_S & = & K^{(+)} (t)\rho_S + \rho_S K^{(-)} (t) + \sum_{\alpha\beta} \int_0^t \mbox{d}u \int_0^t \mbox{d}v 
\, c_{\beta\alpha} (v,u) \, V_{\alpha} (u) \rho_S V_{\beta} (v)\, , \nonumber \\ 
K^{(\eta)} (t) & = &  -\frac{1}{2} \sum_{\alpha\beta} \int_0^t \mbox{d} u  
\int_0^t \mbox{d} v \, \, c^{(\eta)}_{\alpha\beta}(u,v)\, 
{\cal T}_{\eta} \left \{ V_{\alpha}(u) V_{\beta}(v) \right \} \, . 
\label{2}
\end{eqnarray} 
The prescription ${\cal T}$ orders products of system potentials
$\{V_{\alpha}(t)\}_{\alpha}$ according to
\begin{eqnarray} 
{\cal T}\left\{ \prod_{i=1}^m V_{\alpha_i}(t_i) \rho_S \prod_{j=1}^n V_{\alpha'_j}(t'_j)\right\} & = &  
{\cal T}_+ \left\{ \rule{0mm}{7mm} \prod_{i=1}^m  V_{\alpha_i}(t_i) \right\} \rho_S 
{\cal T}_- \left\{ \prod_{j=1}^n V_{\alpha'_j}(t'_j) \right\}\,, \nonumber \\  
{\cal T}_+\left\{ \rule{0mm}{7mm} \prod_{i=1}^m V_{\alpha_i}(t_i)\right\}  & = &   
V_{\alpha_1}(t_1)  \cdots V_{\alpha_m}(t_m)\, , \nonumber \\ 
{\cal T}_-\left\{ \rule{0mm}{7mm} \prod_{j=1}^n V_{\alpha'_j}(t'_j)\right\} & = & 
V_{\alpha'_n}(t'_n) \cdots  V_{\alpha'_1}(t'_1)\, , 
\label{3}
\end{eqnarray} 
where the inequalities $t_1 > \cdots > t_m$ and $t'_1 > \cdots > t'_n$ are assumed. The
dependence on time of the system potentials is determined by the unperturbed system
Hamiltonian $H_S$ as $V_{\alpha} (t) = \exp(iH_St)V_{\alpha} \exp(-iH_St)$.

Since the system Hilbert space is spanned by a countable number of ket vectors, the
interaction with the reservoir can be described by the Hamiltonian
$\sum_{\alpha} V_{\alpha}\otimes U_{\alpha}$. The dummy $\alpha$ takes on a countable
number of values. The reservoir is made up by a continuum of e.m. modes, so the set of
potentials $\{U_{\alpha}\}$ consists of ladder operators. If the reservoir is in a thermal
state $\rho_R$ at time zero, Wick's theorem can be employed to express reservoir
correlation functions of arbitrary order in terms of three pair correlation
functions. These are given by
\begin{eqnarray} 
c_{\alpha_1\alpha_2} (t_1,t_2) & = & \mbox{Tr}_R\left[ U_{\alpha_1}(t_1) U_{\alpha_2}(t_2)\rho_R\right] \, , \nonumber \\
c_{\alpha_1\alpha_2}^{(+)}(t_1,t_2) & = &  c_{\alpha_1\alpha_2} (t_1,t_2) \, \theta (t_1-t_2) + 
c_{\alpha_2\alpha_1} (t_2,t_1) \, \theta (t_2-t_1) \, , 
\nonumber \\  
c_{\alpha_1\alpha_2}^{(-)}(t_1,t_2) & = & c_{\alpha_1\alpha_2} (t_1,t_2) \, \theta (t_2-t_1) + 
c_{\alpha_2\alpha_1} (t_2,t_1) \, \theta (t_1-t_2) \, ,   
\label{4} 
\end{eqnarray} 
where $\theta (t)$ is the Heaviside step function. It should be pointed out that (\ref{1})
is valid if the initial state of system and reservoir can be written as
$\rho_{SR} = \rho_{S} \otimes \rho_{R}$, where $\rho_{R}$ denotes a thermal state.

By working out (\ref{1}) with the help of (\ref{3}) one can expand the density matrix of
$S$ as
\begin{align}
& \rho_S(t) = \sum_{q=0}^{\infty} \sum_{\alpha^{}_1  \cdots \alpha^{}_q} 
\sum_{\alpha'_1  \cdots \alpha'_q} \int_0^t  \mbox{d} t^{}_1  \cdots \int_0^{t^{}_{q-1}}  \mbox{d} t^{}_q  
\int_0^t \mbox{d} t'_1  \cdots  \int_0^{t'_{q-1}}  \mbox{d} t'_q   \nonumber \\[-6mm] \intertext{} 
& \times W_q^{(+)} (t;t^{}_1, \cdots,t^{}_q)_{\alpha^{}_1\cdots\alpha^{}_q} \rho_S 
W_q^{(-)} (t;t'_1,\cdots,t'_q)_{\alpha'_1\cdots\alpha'_q} \, \frac{1}{q!} 
\sum_{P Q} \prod_{k=1}^q 
c_{\alpha'_{Q(k)}\alpha^{}_{P(k)}} (t'_{Q(k)},t^{}_{P(k)}) \, .
\label{5} 
\end{align} 
We sum over all permutations $P$ and $Q$ of the integers $\{1,\ldots,q\}$.  The Kraus
matrices $\{W_q^{(+)}\}_{q\ge 0}$ satisfy the infinite hierarchy
\begin{align} 
& W_0^{(+)} (t)  =  1_S -\sum_{\alpha \beta} \int_0^t \mbox{d} u \int_0^{u} \mbox{d} v \, c_{\alpha \beta}(u,v) \, 
V_{\alpha}(u) W_1^{(+)} (u;v)_{\beta} \, , \nonumber \\[-6mm] \intertext{}  
& W_q^{(+)} (t;t^{}_1, \cdots,t^{}_q)_{\alpha^{}_1\cdots\alpha^{}_q} =  V_{\alpha^{}_1}(t^{}_1) 
W_{q-1}^{(+)} (t^{}_1;t^{}_2,\cdots,t^{}_q)_{\alpha^{}_2\cdots\alpha^{}_q} \nonumber \\[-6mm] \intertext{}  
& -\sum_{j=1}^{q+1}\, \sum_{\alpha \beta}\, \int_{t^{}_1}^t 
\mbox{d} u \int_{t_j}^{t_{j-1}} \mbox{d} v \, c_{\alpha \beta}(u,v) V_{\alpha}(u) 
W_{q+1}^{(+)} (u;t^{}_1,\cdots,t^{}_{j-1},v,t^{}_j,\cdots,t^{}_q)_{\alpha^{}_1\cdots \alpha^{}_{j-1}\beta 
\alpha^{}_j\cdots\alpha^{}_q}\, , 
\label{6}  
\end{align}  
with $t>t_1> \cdots >t_q>0$. In evaluating the boundaries of the integral over $v$ one has
to choose $t_0=u$ and $t_{q+1}=0$. From the solution for $W_q^{(+)}$ and the symmetry
relation
\begin{equation}  
W_q^{(-)} (t;t^{}_1, \cdots,t^{}_q)_{\alpha^{}_1\cdots\alpha^{}_q} = 
\left [ W_q^{(+)} (t;t^{}_1, \cdots,t^{}_q)_{\alpha^{\dagger}_1\cdots\alpha^{\dagger}_q} \right ]^{\dagger}   
\label{7}
\end{equation}
the solution for the Kraus matrices $W_q^{(-)}$ can be found. On the right-hand side of
(\ref{7}) the replacements $V_{\alpha_j}\rightarrow V_{\alpha_j}^{\dagger}$ must be made
for $1\le j \le q$, as indicated by the subscripts $\alpha^{\dagger}_i$.

The material presented in (\ref{1})--(\ref{7}) constitutes a repetition of the treatment
developed in \cite{EPL:2013}.  In this article, we shall solve the Kraus hierarchy
(\ref{6}). To that end, we represent the system potentials by means of the orthonormal
basis of eigenkets $\{|k\rangle\}_{k\ge 1}$ of $H_S$, with eigenvalues
$\{\omega_k \}_{k\ge 1}$.  We thus make the transition
\begin{equation} 
\alpha \rightarrow (kl)\, , \,\,\,\,\,\,\,\,\, V_{\alpha}(t) \rightarrow |k\rangle\langle l| \exp [i\omega_{(kl)}t]\, ,
\label{8}
\end{equation}    
where the shorthand $\omega_{(kl)} = \omega_k - \omega_l$ is used. 

For matrix elements of $W_q^{(+)}$ the notation 
\begin{equation} 
\langle k_1| W_q^{(+)} (t;t^{}_1, \cdots ,t^{}_q)_{(l_1k_2)\cdots (l_qk_{q+1})}|l_{q+1}\rangle = 
W_q(t;t^{}_1, \cdots ,t^{}_q)_{(k_1k_2\cdots k_{q+1})(l_1l_2\cdots l_{q+1})}
\label{9} 
\end{equation}   
will be introduced. Furthermore, the abbreviations 
\begin{align} 
& K^n_q = (k_{n+1}k_{n+2}\cdots k_q)\, , \,\,\,\,  K_q=K^0_q\, , \,\,\,\, T^n_q = t_{n+1}, t_{n+2}, \cdots , t_q\, , \,\,\,\, 
T_q = T^0_q \, , \,\,\,\, \delta_{K_qL_q} = \prod_{s=1}^q \delta_{k_sl_s}\, , \nonumber \\[-6mm] \intertext{} 
& \int_0^t \mbox{d} T^n_q = \int_0^t \mbox{d} t_{n+1} \int_0^{t_{n+1}} \mbox{d} t_{n+2} \cdots \int_0^{t_{q-1}} \mbox{d}t_q\, , 
\,\,\,\,  Z^+_n = z + z_1 + z_2 + \cdots + z_n   
\label{10}
\end{align} 
will be employed. Last, the Laplace transform  
\begin{equation}  
\hspace{-18mm}
\hat{W_q} (z;Z_q)_{K_{q+1}L_{q+1}} = (-i)^{q+1} \int_0^{\infty} \mbox{d}t  \int_0^t \mbox{d} T_q \exp [izt + iZ_q \cdot T_q] 
W_q (t;T_q)_{K_{q+1}L_{q+1}}\, , 
\label{11} 
\end{equation}  
will be invoked, with $\mbox{Im}z>0$ and $\mbox{Im}z_j > 0$ for $1\le j \le q$. 
For $q=0$ the integration over $T_q$ and the variable $Z_q$ must be omitted.   

Transformation of the Kraus hierarchy (\ref{6}) provides us with  
\begin{align} 
& \hat{W}_q (z-\omega_{k_1};Z_q)_{K_{q+1}L_{q+1}} = \delta_{k_1 l_1} (z-\omega_{k_1})^{-1} 
\hat{W}_{q-1} (z + z_1 -\omega_{k_{2}};Z^1_q)_{K^1_{q+1}L^1_{q+1}}  
\label{12} \\[-6mm] \intertext{}  
& - \sum_{j=1}^{q+1} \sum_{klm} \int_C \frac{\mbox{d}y}{2\pi i} (z-\omega_{k_1})^{-1} \hat{c}_{(k_1k)(lm)}(y) 
\hat{W}_{q+1} (z -y -\omega_k ; Z_{j-1}, y, Z^{j-1}_q)_{(kK^1_j m K^j_{q+1})(L_{j-1}lL^{j-1}_{q+1})} \, , 
\nonumber 
\end{align} 
where for $q=0$ the convention $\hat{W}_{-1} = 1$ is in force. The transform $\hat{f}(z)$
is given by $-i\int_0^{\infty} \mbox{d}t \exp[izt] f(t)$, with $f$ any smooth
function. The contour $C$ is parametrized as $-\infty < \mbox{Re}\, y < \infty$, with
$\mbox{Im}\, y$ fixed and $\mbox{Im}\, z> \mbox{Im}\, y>0$.

To get a clue about how (\ref{12}) must be solved we consider a simplified version,
namely, $\hat{W}_q = A \hat{W}_{q-1} + B \hat{W}_{q+1}$, with constant matrices $A$ and
$B$. Since on the right-hand side the index $q$ both increases and decreases, solution by
direct iteration is rather awkward. However, the size of the iterative solution can be
significantly reduced if the matrix ratio $R_q = \hat{W}_q \hat{W}_{q-1}^{-1}$ is
introduced. We then obtain the identity $R_q = [1-BR_{q+1}]^{-1} A$, the iteration of which
yields a matrix continued fraction. As continued fractions frequently occur in quantum 
mechanics, and more specifically quantum optics \cite{SCH:1967}--\cite{SWA:1973}, it seems 
that we have found a natural path for solving the Kraus hierarchy. Even more so because at zero temperature
employment of the ratio $R_q$ in (\ref{12}) reproduces the well-known and exact density
matrix for decay of a two-level atom \cite{LOU:1973}.

At finite temperature, additional preparations have to be made in order to work with
matrix ratios. We have to iterate (\ref{12}) so as to replace it by the more general
hierarchy
\begin{align} 
& \hat{W}_q (z-\omega_{k_1};Z_q)_{K_{q+1}L_{q+1}} = \delta_{K_n L_n} \prod_{s=1}^n (Z^+_{s-1}-\omega_{k_s})^{-1} 
\hat{W}_{q-n} (Z^+_n -\omega_{k_{n+1}};Z^n_q)_{K^n_{q+1}L^n_{q+1}} \nonumber \\[-6mm] \intertext{} 
& -\sum_{p=-1}^{n-2} \sum_{j=p+2}^{q+1} \sum_{klm} \int_C \frac{\mbox{d}y}{2\pi i}\,  \delta_{K_{p+1}L_{p+1}} 
\prod_{s=1}^{p+2} (Z^+_{s-1}-\omega_{k_s})^{-1} \hat{c}_{(k_{p+2}k)(lm)}(y) \nonumber \\[-6mm] \intertext{}  
& \times \hat{W}_{q-p} (Z^+_{p+1} -y -\omega_k ; Z^{p+1}_{j-1}, y, 
Z^{j-1}_q)_{(kK^{p+2}_j m K^j_{q+1})(L^{p+1}_{j-1}lL^{j-1}_{q+1})} \, ,  
\label{13} 
\end{align}  
with the conditions $q\ge 0$ and $0\le n \le q+1$. Now it can be recognized that a set of
matrix ratios must be defined according to
\begin{equation}  
\hspace{3mm}
R_{q,n} (z;Z_q)_{K_{q+1}L_{q+1}} = \sum_{M_{q+1}} \hat{W}_q (z-\omega_{k_1}; Z_q)_{K_{q+1}M_{q+1}} \delta_{M_nL_n} 
\hat{W}_{q-n}^{-1} (Z^+_n - \omega_{m_{n+1}}; Z^n_q)_{M^n_{q+1}L^n_{q+1}}\, , 
\label{14} 
\end{equation} 
for $0\le n \le q+1$. On the right-hand side a matrix inverse is taken, so one has 
$R_{q,0} = \delta_{K_{q+1}L_{q+1}}$. 

On the left-hand side of (\ref{13}) the unit matrix appears upon multiplying from the
right by $\hat{W}_q^{-1} (z-\omega_{k_1};Z_q)_{K_{q+1}L_{q+1}}$. On the right-hand side we
insert the identity $I^{-1}I = 1$ in order to get a closed set of equations in terms of
matrix ratios. The intermediate matrix must be chosen as
\begin{equation}  
I(z;Z_q)_{K_{q+1}L_{q+1}} = \delta_{K_jL_j} \hat{W}_{q-j} (Z^+_j - \omega_{k_{j+1}}; Z^j_q)_{K^j_{q+1}L^j_{q+1}} \, .
\label{15} 
\end{equation}  
We then find 
\[  
 R_{q,n}^{-1} (z; Z_q)_{K_{q+1}L_{q+1}} =  \delta_{K_{q+1}L_{q+1}} \prod_{s=1}^n (Z^+_{s-1} - \omega_{k_s}) 
+ \sum_{p=-1}^{n-2} \sum_{j=p+2}^{q+1} \sum_{M_{q+1}} \sum_{klm} \int_C \frac{\mbox{d}y}{2\pi i}  \delta_{K_{p+1}M_{p+1}} 
\]    
\[  
\hspace{-16mm} 
\times  \prod_{s=p+3}^n (Z^+_{s-1} - \omega_{k_s}) 
R_{q-p, j-p} (Z^+_{p+1}-y; Z^{p+1}_{j-1},y,Z^{j-1}_q)_{(kK^{p+2}_jmK^j_{q+1})(M^{p+1}_{j-1}lM^{j-1}_{q+1})} 
\] 
\begin{equation}  
\hspace{-10mm}  
\times  R^{-1}_{q,j}(z;Z_q)_{M_{q+1}L_{q+1}} \hat{c}_{(k_{p+2}k)(lm)}(y)    \, ,
\label{16}
\end{equation} 
with $q\ge 0$ and $0\le n \le q+1$. The boundaries of the summations over $p$ and $j$
guarantee that both $R_{q-p,j-p}$ and $R_{q,j}$ satisfy the afore-mentioned
conditions, so indeed we have a closed set in our hands.  Its solution directly produces
the Kraus matrices, in view of the property
$R_{q,q+1} (z;Z_q)_{K_{q+1}L_{q+1}} = \hat{W}_q (z-\omega_{k_1};Z_q)_{K_{q+1}L_{q+1}}$.

\section{Continued fractions}  

In the previous section, we have replaced the Kraus hierarchy (\ref{6}) by the closed set
(\ref{16}) for matrix ratios.  Construction of the iterative solution of (\ref{16}) goes
in two steps: we start by eliminating $R^{-1}$ on the right-hand side of (\ref{16}). Then,
in the ensuing equation we perform a finite interation in $R$.

From (\ref{16}) we obtain an expression for
$R^{-1}_{q,j}(z;Z_q)_{M_{q+1}L_{q+1}}$ that is substituted on the right-hand side. 
Continuation of this process ad infinitum brings us to
\begin{align} 
& R^{-1}_{q,j_0}(z;Z_q)_{K_{q+1,1}K_{q+1,2}} =  \prod_{s=1}^{j_0} (Z^+_{s-1} - \omega[s,1]) 
\prod_{s=1}^{q+1} \delta(s,1;s,2)  \nonumber \\ 
& + \sum_{h_1=1}^{\infty}\sum_{p_1=-1}^{j_0-2}\sum_{j_1=p_1+2}^{q+1}\cdots 
\sum_{p(h_1)=-1}^{j(h_1-1)-2}\sum_{j(h_1)=p(h_1)+2}^{q+1} 
\underset{{\displaystyle \{k_{a,b}\}_{a=1\,\,\, b=3}^{q+1\,\,\, h_1+3}}}{\sum} \, 
\underset{{\displaystyle \{k_{q+2,b+1},k_{q+3,b},k_{q+4,b}\}_{b=3}^{h_1+2}}}{\sum} \, \nonumber \\ 
& \times \int_C \frac{\mbox{d} z_{q+1}}{2\pi i} \cdots \int_C \frac{\mbox{d} z_{q+h_1}}{2\pi i} 
\prod_{s=1}^{q+1} \left [\delta (s,1;s,3) \delta (s,h_1+3;s,2)\right ]  
\prod_{s=1}^{j(h_1)} (Z^+_{s-1} - \omega[s,h_1+3])  \nonumber \\
& \times \prod_{m_1=1}^{h_1} \left [  \rule{0mm}{9mm}
\prod_{s=1}^{p(m_1)+1} \delta (s,m_1+2;s,m_1+3) \prod_{s=p(m_1)+3}^{j(m_1-1)} 
(Z^+_{s-1} - \omega[s,m_1+2]) \right.  \nonumber \\
& \times \left \{ p_{m_1}+2,m_1+2;q+3,m_1+2;q+2,m_1+3;q+4,m_1+2 \right \}_{N=0}  \nonumber \\ 
& \times R_{q-p(m_1),j(m_1)-p(m_1)}(Z^+_{p(m_1)+1}-z_{q+m_1}; 
Z^{p(m_1)+1}_{j(m_1)-1},z_{q+m_1},Z^{j(m_1)-1}_q)_{(L')(L'')} 
\left.  \rule{0mm}{9mm} \right ]  \, . \label{17} 
\end{align}
We used the following notations:
\begin{align} 
&  K_{q,a} = (k_{1,a}k_{2,a}\cdots k_{q,a})\, , \nonumber \\ 
&  \omega[a,b]=\omega_{k_{a,b}}\, , \,\,\,\,   
\delta(a,b;c,d) = \delta_{k_{a,b}k_{c,d}} \, , 
\,\,\,\, 
m(n) = m_n \, , \,\,\,\, 
\nonumber \\
& \left \{ a,b;c,d;e,f;g,h \right \}_N = \hat{c}_{(k_{a,b}k_{c,d})(k_{e,f}k_{g,h})}(z_{q+m_{N+1}})\, .
\label{18}
\end{align} 
Furthermore, the indices $L'$ and $L''$ must be replaced by 
\begin{eqnarray}
L' & = &  k_{q+3,m_1+2}k_{p(m_1)+3,m_1+2}\cdots 
k_{j(m_1),m_1+2}k_{q+4,m_1+2}k_{j(m_1)+1,m_1+2}\cdots k_{q+1,m_1+2}\, ,
\nonumber \\ 
L''& = & k_{p(m_1)+2,m_1+3}\cdots 
k_{j(m_1)-1,m_1+3}k_{q+2,m_1+3}k_{j(m_1),m_1+3}\cdots k_{q+1,m_1+3} \, .
\label{19}
\end{eqnarray} 
Identity (\ref{17}) marks the completion of step one. 

Step two is much harder and consists of performing for (\ref{17}) a finite number of
iterations in $R_{q,j_0}$.  In order to cast (\ref{17}) into a form that indeed allows for
iteration, we define a number of permutations. The permutations $C_{j,q}$ and $D_{N,q}$ are
given by
\begin{align} 
C_{j,q}(i) = \begin{cases}
i     & \text{if $i \le j-1$}\,, \\
q+1   & \text{if $i = j$}\,,     \\
i-1   & \text{if $j + 1 \le i \le q + 1$}\,,  \\
i     & \text{if $q + 2 \le i$}\,, 
\end{cases} 
 & & 
D_{N,q} (i) = \begin{cases}
i     & \text{if $i \le p_{m_N}+2$}\,, \\
q+3   & \text{if $i = p_{m_N} + 3$}\,,     \\
i-1   & \text{if $p_{m_N} + 4 \le i \le j_{m_N} + 1$}\,,  \\
q+4   & \text{if $i = j_{m_N} + 2$}\,,     \\
i-2   & \text{if $j_{m_N} + 3 \le i \le q + 4$}\,,  \\
i     & \text{if $q + 5 \le i$}\,, 
\end{cases} 
\nonumber  \\[-6.5mm] & & \label{20} 
\end{align} 
with $N=1,2,3,\ldots$. We furthermore define 
\begin{equation} 
E_{N,q}(i) = C_{j(m(N)),q} (i) \, ,  
\label{21} 
\end{equation}   
with $N=1,2,3,\ldots$. 
For $N=0$ the definitions $D_{0,q} (i) = i-2$ if $3 \le i \le q+3$
and $E_{0,q} (i) = i-1$ if $2 \le i \le q+2$ will be employed. The permutation $F_{N,q}$
must be constructed from
\begin{equation} 
F_{N,q}(i) = C_{j(m(1)),q+m(1)-1} \left ( C_{j(m(2))+1,q+m(2)-1} \left ( \cdots 
C_{j(m(N))+N-1,q+m(N)-1}(i)\cdots \right) \right) \, .     
\label{22} 
\end{equation}  
For $N=0$ one has $F_{0,q}(i)=i$ for all integers $i$. 

Furthermore, we introduce some rules for writing up long expressions in a concise
manner. We shall make use of the following notation:
\begin{align} 
& \bar{h}_N = h_N + N + 1\, , \,\,\,\,\,\, \bar{m}_N = m_N + N + 1\, ,  \nonumber \\
& Z^{+-}_{n,N} = z + \sum_{i=1}^n z_{F_{N,q}(i)} - \sum_{i=1}^N z_{q+m_i}\, , \,\,\,\,\,\, 
\underset{\!\!\!N}{\mathrlap{\int}\raisebox{0.5ex}{\rule{4mm}{0.3mm}}}  \, = 
[1/(2\pi i)]^{h_{N+1}-m_N} \int_C \mbox{d} z_{q+m_N+1} \cdots \int_C \mbox{d} z_{q+h_{N+1}} \, , 
\nonumber \\[-6mm] \intertext{} 
& q'_N = q - p(m_{N+1})\, , \,\,\,\,\,\, j'_N = j(m_{N+1}) - p(m_{N+1}) \, , \,\,\,\,\,\, 
z'_N = Z_{p(m_{N+1})+N+1,N+1}^{+-} \, ,
\nonumber \\[-6mm] \intertext{} 
& K'_N = (k_{D_{N+1,q}(p(m_{N+1})+3),m_{N+1}+N+2}\, k_{D_{N+1,q}(p(m_{N+1})+4),m_{N+1}+N+2}\ldots 
k_{D_{N+1,q}(q+3),m_{N+1}+N+2}) \, ,  
\nonumber \\[-6mm] \intertext{} 
& K''_N = (k_{E_{N+1,q+1}(p(m_{N+1})+2),m_{N+1}+N+3}\, k_{E_{N+1,q+1}(p(m_{N+1})+3),m_{N+1}+N+3} \ldots 
k_{E_{N+1,q+1}(q+2),m_{N+1}+N+3}) \, ,  
\nonumber \\[-6mm] \intertext{} 
& Z''_N = z_{F_{N+1,q}(p(m_{N+1})+N+2)}, z_{F_{N+1,q}(p(m_{N+1})+N+3)}, \ldots, z_{F_{N+1,q}(q+N+1)}\, ,  
\label{23} 
\end{align} 
where the permutations $D_{N,q}, E_{N,q},$ and $F_{N,q}$ are specified in
(\ref{20})--(\ref{22}). 

As a final abbreviation we introduce the symbol
$\mathrlap{\sum_N}\raisebox{0.4ex}{\rule{4mm}{0.3mm}}\,\,\,\, $ to indicate a summation
  over integers $h_{N+1}\,$, $\{j_i, p_i \}_{i=m_N+1}^{h_{N+1}}\,$ and
  $\{k_{a,b}\}_{a=p_{m_N}+1}^{q+4}\, \mbox{}_{b=\bar{m}_N+2}^{\bar{h}_{N+1}+1}$, except
  for $\{k_{q+2,\bar{m}_N+2}\, , k_{q+3,\bar{h}_{N+1}+1}\, ,k_{q+4,\bar{h}_{N+1}+1} \}$,
  with boundaries given by
\begin{align}
& m_N+1 \le h_{N+1}< \infty\, ,  \,\,\,\,\, 
\{ p_{m_N}-1 \le p_i \le j_{i-1}-2\, , p_i+2\le j_i \le q+1 \}_{i=m_N+1}^{h_{N+1}}    \, . 
\label{24}
\end{align}
In the summation with horizontal bar each integer
$k_{a,b}$ runs over the same values as the label $k$ of the eigenket $|k\rangle$ of
$H_S$. It should be stressed that the prescriptions for sum and integral with horizontal
bar should be transferred in facsimile to formulas that are completely written out. The
only component that may be adapted is integer $N$, which may take on any
nonnegative value. Last, the special choices $m_0 = p_0=0$ are in force.

With the above instruments in hand we can shorten our continued fractions
considerably. Instead of (\ref{17}) we may write
\begin{equation} 
R^{-1}_{q,j_0}(z;Z_q)_{K_{q+1,1}K_{q+1,2}} =  A_0 + B_0 R_{q'_0,j'_0}(z'_0; Z''_0)_{K'_0K''_0}\, , 
\label{25} 
\end{equation}  
where the prescription for the new components reads  
\begin{eqnarray} 
A_N \!\!&\!\! =\!\! &\!\! \left [ \prod_{s=p(m_N)+3}^{q+3} \delta (D_{N,q}(s), \bar{m}_N; E_{N,q+1}(s-1), \bar{m}_N+1) \right ] \!\!\! 
          \left [ \prod_{s=p(m_N)+3}^{j(m_N)+2} (Z_{s+N-3,N}^{+-} - \omega [D_{N,q} (s), \bar{m}_N] ) \right ], \nonumber \\ 
B_N \!\!&\!\! =\!\! &\!\!  \mathrlap{\sum_N}\raisebox{0.4ex}{\rule{4mm}{0.3mm}}\,\,\,\,\,
          \underset{\!\!\!N}{\mathrlap{\int}\raisebox{0.5ex}{\rule{4mm}{0.3mm}}}  \!
          \left [ \prod_{s=p(m_N)+3}^{q+3} \!\!\! \delta (D_{N,q}(s), \bar{m}_N; s-2, \bar{m}_N+2) 
          \delta (s-2, \bar{h}_{N+1}+1; E_{N,q+1}(s-1), \bar{m}_N+1) \right ]  \nonumber \\ 
    &   & \times \left [ \prod_{s=p(m_N)+1}^{j(h_{N+1})} (Z_{s+N-1,N}^{+-}-\omega[s,\bar{h}_{N+1}+1]) \right ] \nonumber \\ 
    &   & \times \prod_{m_{N+1}=m_N+1}^{h_{N+1}}\left \{ p_{m_{N+1}}+2,\bar{m}_{N+1};q+3,\bar{m}_{N+1}; 
           q+2,\bar{m}_{N+1}+1;q+4,\bar{m}_{N+1} \right \}_N \nonumber \\ 
    &   & \times \left [ \prod_{s=p(m_N)+1}^{p(m_{N+1})+1} \delta (s,\bar{m}_{N+1};s,\bar{m}_{N+1}+1) \right ] \!\!\! 
          \left [ \prod_{s=p(m_{N+1})+3}^{j(m_{N+1}-1)} (Z_{s+N-1,N}^{+-} - \omega [s,\bar{m}_{N+1}]) \right ] \, .  
\label{26}  
\end{eqnarray} 
Again, these components must be implemented in facsimile, the integer $N\ge 0$ being the
only variable that may be modified. 

Iteration of (\ref{25}) gives 
\begin{align} 
& R_{q,j_0}(z;Z_q)_{K_{q+1,1}K_{q+1,2}}  =   
\prescript{}{}{\left \{ \rule{0mm}{5mm} \right. } A_0 + B_0 
\prescript{}{}{\left \{ \rule{0mm}{5mm} \right. } A_1 + B_1 
\prescript{}{}{\left \{ \rule{0mm}{5mm} \right. } A_2 + \cdots  \nonumber \\[-6mm] \intertext{}  
& \cdots + B_{N-1}\prescript{}{}{\left \{ \rule{0mm}{5mm} \right. } A_N + B_N 
R_{q'_N,j'_N}(z'_N;Z''_N)_{K'_NK''_N}  
\left. \rule{0mm}{5mm} \right \}_{}^{-1}      
\cdots \left. \rule{0mm}{5mm} \right \}_{}^{-1}
\left. \rule{0mm}{5mm} \right \}_{}^{-1}
\left. \rule{0mm}{5mm} \right \}_{}^{-1} \, ,
\label{27} 
\end{align} 
with $N=0,1,2,\ldots$. The superscript $-1$ of each right bracket indicates that a 
{\em matrix} inverse must be taken. 

From (\ref{26}) it follows that on the right-hand side of (\ref{25}) a product 
over $m_1$ must be carried out. We stress that this product over $m_1$ pertains to the full expression 
$B_0 R$, where indices and arguments of the matrix $R$ have been omitted. For the sum 
$\mathrlap{\sum_{N=0}}\raisebox{0.4ex}{\rule{4mm}{0.3mm}}$ $\,\,\,\,\,\,\,\,$ and integral 
${\mathrlap{\int}\raisebox{0.5ex}{\rule{4mm}{0.3mm}}}_{N=0}$ 
figuring on the right-hand side of (\ref{25}) the same remark applies. Consequently, iteration of 
(\ref{25}) produces in (\ref{27}) a most complicated analytic structure. 

The proof of (\ref{27}) is based on induction in $N$. Upon substituting the relation 
\begin{equation} 
R_{q'_N,j'_N}(z'_N;Z''_N)_{K'_NK''_N}  = 
\left \{ A_{N+1} + B_{N+1} R_{q'_{N+1},j'_{N+1}}(z'_{N+1}; Z''_{N+1})_{K'_{N+1}K''_{N+1}}  
\right \}^{-1}\, 
\label{28} 
\end{equation}  
into (\ref{27}), we indeed reproduce (\ref{27}) with the replacement $N\rightarrow N+1$
made.  Therefore, it is sufficient to verify that (\ref{28}) holds true. This can be shown
by carrying out a suitable set of consecutive transformations of the variables and labels
in (\ref{17}). Last, by choosing $N=-1$ in (\ref{28}) we reproduce (\ref{25}), as required.  
We thus have demonstrated that a finite iteration of (\ref{25}) provides us with (\ref{27}).

\section{Density matrix} 

If the conditions $\mbox{Im}(z+z^{}_1+\cdots + z^{}_j)> \gamma_j$ are fulfilled for 
$0 \le j \le q$, then all matrix ratios $R_{q,j_0} (z;Z_q)$ are analytic in $z$ and 
$z_1$, $z_2$, \ldots , $z_q$. Hence, it is reasonable to assume that the matrix continued 
fraction (\ref{27}) converges for $N\rightarrow \infty$ as long as the inequalities 
$\mbox{Im}(z+z^{}_1+\cdots + z^{}_j)> \gamma_j$ are satisfied for $0 \le j \le q$. The
fixed numbers $\{\gamma_j\}_{j=0}^q$ depend on such parameters as the coupling constant for 
the interaction between system and thermal reservoir; in this work the set 
$\{\gamma_j\}_{j=0}^q$ will not be specified any further.

Convergence of (\ref{27}) implies that we may set $R_{q'_N,j'_N}$ equal to zero for large
$N$ and the inequalities $\mbox{Im}(z+z^{}_1+\cdots + z^{}_j)> \gamma_j$ true, with 
$0 \le j \le q$. The solution for the Kraus matrices is thus found as
\begin{align} 
& \hat{W}_q (z - \omega_{k_{1,1}};Z_q)_{K_{q+1,1}K_{q+1,2}} = 
R_{q,q+1}(z;Z_q)_{K_{q+1,1}K_{q+1,2}} = \nonumber   \\[-6mm] \intertext{} 
& \lim_{N\rightarrow \infty}\,\, \prescript{}{}{\left \{ \rule{0mm}{5mm} \right. } A_0 + B_0 
\prescript{}{}{\left \{ \rule{0mm}{5mm} \right. } A_1 + B_1 
\prescript{}{}{\left \{ \rule{0mm}{5mm} \right. } A_2 +   
 \cdots + B_{N-1}\prescript{}{}{\left \{ \rule{0mm}{5mm} \right. } A_N   
\left. \rule{0mm}{5mm} \right \}_{}^{-1}      
\cdots \left. \rule{0mm}{5mm} \right \}_{}^{-1}
\left. \rule{0mm}{5mm} \right \}_{}^{-1}
\left. \rule{0mm}{5mm} \right \}_{}^{-1}\rule[-4mm]{0.1mm}{10mm}_{\, j_0=q+1} \,\,\, . 
\label{29}
\end{align}  
The components $A_N$ and $B_N$, defined in (\ref{26}), must be implemented in facsimile. 

The solution for the Kraus operators being available, we can return to the density matrix
as given by (\ref{5}). In order to express (\ref{5}) in terms of Laplace transforms
throughout, we substitute the inverse of the transformation (\ref{11}).  Next, the Laplace
representation of the correlation functions can be found with the help of the identity
$c_{\alpha'\alpha}(t',t) = c_{\alpha'\alpha}(t'-t,0)$ and the relations
\begin{eqnarray} 
c_{\alpha'\alpha}(t,0) & = & \frac{i}{2\pi} \int_{-\infty}^{\infty} \mbox{d}y \exp (-iyt)  
\mathrlap{\hat{c}}\raisebox{0.3ex}{\hspace{0.27em}$\hat{}$}_{\,\,\alpha'\alpha}(y) \, , \nonumber \\ 
\mathrlap{\hat{c}}\raisebox{0.3ex}{\hspace{0.27em}$\hat{}$}_{\,\,\alpha'\alpha}(y)  & = & 
\hat{c}_{\alpha'\alpha , +}(y)+\hat{c}_{\alpha'\alpha , -}(y) \, ,  \nonumber \\ 
\hat{c}_{\alpha'\alpha , \eta} (y) & = & -i \int_0^{\infty} \mbox{d} t \exp (i\eta yt) c_{\alpha'\alpha}(\eta t,0)\, ,  
\label{30} 
\end{eqnarray} 
with $\eta = \pm 1$. Note that the difference $t'-t$ can become negative, so we do need
the backward transform $\hat{c}_{\alpha'\alpha , -} (y)$.

Once we have switched to Laplace representation, the temporal integrals of (\ref{5})
become elementary. We must evaluate the repeated integral
\begin{equation} 
J_q(t;Z_q) = (-i)^q \int_0^t \mbox{d} T_q \exp(iZ_q\cdot T_q) \, ,
\label{31}
\end{equation}  
with $q=1,2,3,\ldots$. The answer reads  
\begin{equation} 
J_q(t;Z_q) = \sum_{p=0}^q (-1)^p \exp [i(Z^+_p-z)t] \prod_{k=1}^p (Z^+_p-Z^+_{k-1})^{-1} 
\prod_{k=p+1}^q (Z^+_k-Z^+_p)^{-1} \,  , 
\label{32}
\end{equation} 
a result that can be proved by means of induction in $q$. Note that in (\ref{5}) time
arguments of correlation functions are paired in all possible ways by the permutations $P$
and $Q$ of the integers $\{1,2,\ldots, q\}$. All of these pairings can be transferred to
the Laplace variables via the identity
\begin{equation} 
\sum_{j=1}^q  z_j t_{P(j)} = \sum_{j=1}^q  z_{P^{-1}(j)} t_{j}\, ,
\label{33}
\end{equation}   
where of course $P$ may be replaced by $Q$. 

With the foregoing technical preparations completed we can present the elements of the
exact density matrix as
\begin{align} 
& \langle k^{}_1 | \rho_S (t) | k'_1 \rangle = \sum_{q=0}^{\infty}  \sum_{PQ} 
\sum_{K^1_{q+1} L^{}_{q+1} K'^1_{q+1} L'_{q+1}} \int_{C^{}_0} \frac{\mbox{d}z^{}_0}{2\pi} \int_{C^{}_1} \frac{\mbox{d}z^{}_1}{2\pi} 
\cdots \int_{C^{}_q} \frac{\mbox{d}z^{}_q}{2\pi} \int_{C_0^{\ast}} \frac{\mbox{d}z'_0}{2\pi} \int_{C_1^{\ast}} \frac{\mbox{d}z'_1}{2\pi} 
\cdots \int_{C_q^{\ast}} \frac{\mbox{d}z'_q}{2\pi}  \nonumber \\[-3mm] \intertext{} 
& \times \int_{-\infty}^{\infty} \frac{\mbox{d}y_1}{2\pi i} \cdots \int_{-\infty}^{\infty} \frac{\mbox{d}y_q}{2\pi i} 
\exp[-iz^{}_0t +iz'_0t]J_q(t;\{y_{P^{-1}(j)}-z^{}_j\}_{j=1}^q) J_q(t;\{-y_{Q^{-1}(j)}+z'_j\}_{j=1}^q) \nonumber \\[-2mm] \intertext{}  
& \times \hat{W}_q(z^{}_0;Z^{}_q)_{K^{}_{q+1}L^{}_{q+1}} \langle l^{}_{q+1} | \rho_S | l'_{q+1}\rangle 
[\hat{W}_q(z'^{\,\ast}_0;Z'^{\,\ast}_q)_{L'_{q+1}K'_{q+1}}]^{\dagger} \nonumber \\[-6mm] \intertext{}  
& \times \frac{1}{q!} \prod_{j=1}^q 
\mathrlap{\hat{c}}\raisebox{0.3ex}{\hspace{0.27em}$\hat{}$}_{{\,\displaystyle (k'_{Q(j)+1}l'_{Q(j)})(l^{}_{P(j)}k^{}_{P(j)+1})}}(y_j)\, . 
\label{34} 
\end{align} 
The contours $\{C^{}_j, C^{\ast}_j\}_{j=0}^q$ run parallel to the real axis from $-\infty$
to $+\infty$, obeying the conditions
$\mbox{Im}(z^{}_0+z^{}_1+\cdots + z^{}_j)> \gamma_j$ and
$\mbox{Im}(z'_0+z'_1+\cdots + z'_j)< - \gamma_j$ for
$0 \le j \le q$. The set $\{\gamma_j \}_{j=0}^q$ is made up of fixed positive numbers. The
adjoint Kraus operator must be found from
\begin{equation}  
[\hat{W}_q(z^{\,\ast}_0;Z^{\,\ast}_q)_{L_{q+1}K_{q+1}}]^{\dagger} = 
[\hat{W}_q(z^{\,\ast}_0;Z^{\,\ast}_q)_{K_{q+1}L_{q+1}}]^{\ast}\, . 
\label{35} 
\end{equation} 
For the Kraus operators one must substitute the solution (\ref{29}) in terms of matrix
continued fractions.
 
The result (\ref{34}) can be seen as the solution of the time-ordering problem posed by
(\ref{1}). In its turn, the concise representation (\ref{1}) relies on the possibility of
factorizing reservoir correlation functions with the help of Wick's theorem.  A reservoir
for which Wick's theorem indeed holds true is given by a continuum of harmonic oscillators
at temperature $\beta^{-1}$.  For such a setting the reservoir potentials come out as
\begin{equation} 
U_{(kl)} = \int_0^{\infty} \mbox{d} \omega \, \lambda (\omega )_{(kl)} a(\omega ) + 
\int_0^{\infty} \mbox{d} \omega \, \lambda (\omega )_{(lk)}^{\ast} a^{\dagger }(\omega ) \, ,  
\label{36}
\end{equation}  
where $a(\omega )$ and $a^{\dagger}(\omega )$ are the ladder operators of the mode of
frequency $\omega $. By $ \lambda (\omega )_{(kl)}$ we denote the coupling constant for
the transition $|l\rangle \rightarrow |k\rangle $ within $S$ as induced by the reservoir
mode of frequency $\omega $. Starting from (\ref{36}) and adopting the Laplace
representation (\ref{30}), we derive for the reservoir correlation functions the
expressions
\begin{eqnarray} 
\mathrlap{\hat{c}}\raisebox{0.3ex}{\hspace{0.27em}$\hat{}$}_{\,\,(kl)(mn)}(y) & = & 
2\pi i \int_0^{\infty} \mbox{d} \omega \,
\lambda (\omega)_{(kl)} \lambda (\omega)_{(nm)}^{\ast} (\mbox{e}^{-\beta\omega}-1)^{-1} \delta (y-\omega) \nonumber \\ 
 & & 
+ \,\, 2\pi i \int_0^{\infty} \mbox{d} \omega \,  
\lambda (\omega)_{(lk)}^{\ast} \lambda (\omega)_{(mn)} (1-\mbox{e}^{\beta\omega})^{-1} \delta (y+\omega)
\, , \label{37} 
\end{eqnarray} 
with $\delta (\omega)$ the Dirac delta function. Therefore, after insertion of (\ref{37})
into (\ref{34}) we can exchange the integrals over $\{y_j\}_{j=1}^q$ for integrals over
mode frequencies.

\section{Conclusion}      

In this work, we have obtained the Kraus map for the density operator of a quantum system that exchanges 
energy with a large reservoir. The assumptions on which our result relies have been described in the 
Introduction.  

The Kraus matrices making up the Kraus map possess a most complex analytic structure, determined by matrix 
continued fractions of a very sophisticated form. Moreover, in order to compute the density operator of the 
open quantum system one must still carry out an infinite sum over the set of Kraus matrices. Therefore, one 
may seriously doubt the existence of an exact master equation for the density operator itself. This last 
judgment has also been put forward in the literature \cite{CHR:2013}--\cite{FER:2017}. 

In summary, we conclude that the system-reservoir formalism provides us with a viable \linebreak description 
of quantum dissipative processes as long as the coupling between system and \linebreak reservoir is weak. 
In that case, the memory time of the reservoir is much shorter than the time scale on which the system 
evolves. Consequently, a perturbation theory can be set up in which Kraus matrices are factorized from a 
certain order onwards. For the perturbed density matrix of the system conservation of trace and positivity 
can be explicitly proved \cite{JPA:2018}. 

In contrast, if one refrains from making any approximations, then the dissipative evolution of the 
system must be found from (\ref{34}). In view of the fact that the structure of this equation is extremely  
difficult, the system-reservoir approach seems to be unsuited if it comes to performing practical 
computations for a system that is strongly coupled to a reservoir.

}
\end{document}